%% file: main.tex
\documentclass[11pt,a4paper]{article}
\usepackage[T1]{fontenc}
\usepackage[utf8]{inputenc}
\usepackage{lmodern}
\usepackage[margin=16mm,top=20mm,bottom=20mm]{geometry}
\usepackage{amsmath,amssymb,graphicx}
\usepackage[numbers,sort&compress]{natbib}
\usepackage{placeins,flafter}
\usepackage[hidelinks]{hyperref}
\newcommand{\avg}[1]{\left\langle #1\right\rangle}
\newcommand{\Ehaar}{\mathbb{E}_{\mathrm H}}
\hypersetup{pdftitle={Erasing and Regenerating the Turbulent Cascade Arrow},pdfauthor={Yang Zhao, Daocheng Zhou, Hongfu Zhang}}
\begin{document}
\begin{center}
{\Large\bfseries Erasing and Regenerating the Turbulent Cascade Arrow}
\end{center}
\input{authors.tex}
\begin{abstract}
Which information tells a turbulent flow which way to cascade? We apply Haar-distributed Hermitian phases that retain prescribed same-mode quadratic information, including modal amplitudes, the energy spectrum, and same-mode two-time correlations. For the phase-Haar ensemble, the expectation of any integrable odd flux functional is zero. Finite ensembles strongly suppress mean transfer and forward-event bias in the primary block. An independent block retains sign separation but fails the scale-8 mean criterion. In matched evolution protocols, nonlinear Navier--Stokes dynamics restores the selected signed-flux scalar close to its natural reference, whereas linear evolution and an amplitude-matched frozen control do not. Flux and a local cubic geometry scalar recover together, yet high local correlation can survive erasure of the arrow. Signed transfer thus requires information beyond the retained quadratic record; the tested nonlinear dynamics rebuilds it.
\end{abstract}

An energy spectrum tells us how much turbulent energy occupies each scale. It does not tell us which way that energy is moving.

In three dimensions, the mean cascade is downscale even though local forward and backward events coexist \citep{r1,r2,r3,r4,r5,r6,r11,r16}. Coarse-grained flux makes the distinction explicit: the energy stored in a scale band and the transfer through that band are different quantities \citep{r7,r8,r9,r10,r12,r13}. Here we call the signed part of interscale transfer its \emph{cascade arrow}. Mean flux, forward-event bias, and finite-window accumulation provide complementary ways to measure it.

Fourier phases can carry information missing from the spectrum. Phase alignment changes exchange in Burgers turbulence and related phase-coupling models \citep{r28,r30,r31,r32}; dynamical phases also organize transfer in wave systems and shell models \citep{r35,r36,r43,r34}. In hydrodynamic turbulence, triad geometry and helicity constrain the possible exchanges \citep{r37,r38,r39,r40,r41}, and direct phase manipulations reveal their importance in decaying flow \citep{r33}. These results establish that phases matter. The unresolved question here is what signed-transfer information can change while an explicitly specified quadratic record remains fixed. Retaining only a spectrum would leave many possible explanations open. Retaining the same-mode tensor and its two-time record imposes a sharper constraint. A field can have the correct spectrum at every instant and still carry a different modal history: multitime correlations encode information beyond a sequence of equal-time spectra \citep{r26,r27}.

We therefore use one scalar phase per mode, shared across velocity components and all sampled times. The intervention connects phase-based surrogate methods \citep{r45,r46,r48} to a symmetry question: can a prescribed quadratic record coexist with opposite transfer directions? We first establish that it can, then ask how the original dynamics rebuilds the lost signed signal. The arrow concerns interscale transfer, distinct from trajectory irreversibility \citep{r17,r18} but related to cascade irreversibility and transfer fluctuations \citep{r20,r21}.

\begin{figure}[!htbp]
\centering\includegraphics[width=\textwidth]{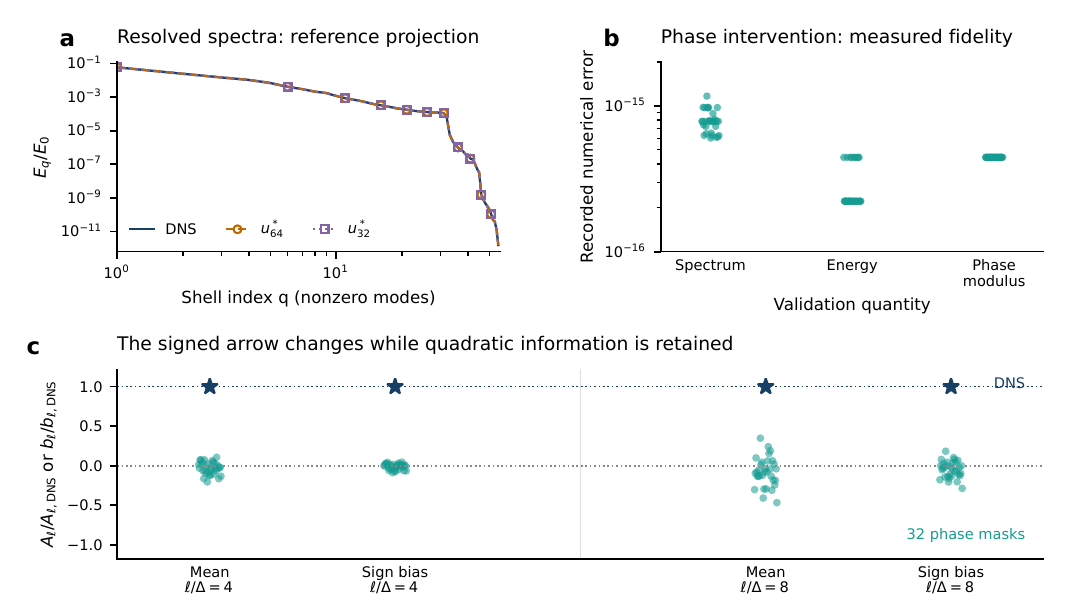}
\caption{Phase intervention and the information boundary. (a) DNS and projected-reference shell spectra with a common DNS energy normalization $E_0=\sum_{q\geq0}\overline E_q^{\rm DNS}$. Only nonzero recorded shells $q$ are plotted; the zero shell is included in $E_0$. (b) Spectrum, energy, and phase-modulus errors for the 32 primary masks. (c) The normalized mean $A_\ell=\avg{\Pi_\ell}/\sigma(\Pi_\ell)$ and instantaneous bias $b_\ell=P(\Pi_\ell>0)-1/2$, each divided by its DNS counterpart and grouped by filter scale. Each turquoise point in (b,c) is one mask; the narrow horizontal offsets only separate points within a category. The quadratic record in Eq.~\eqref{eq:quadratic} is retained while the signed signal is strongly suppressed. Panel (c) is a compact summary of the primary data expanded in Fig.~\ref{fig:static}.}
\label{fig:intervention}
\end{figure}

For each nonzero wave vector, the intervention is
\begin{equation}
 \widehat{\boldsymbol u}^{\phi}(\boldsymbol k,t)
 =\zeta_{\boldsymbol k}\widehat{\boldsymbol u}(\boldsymbol k,t),
 \qquad |\zeta_{\boldsymbol k}|=1,
 \qquad \zeta_{-\boldsymbol k}=\zeta_{\boldsymbol k}^{*}.
 \label{eq:phase}
\end{equation}
The zero mode is fixed. Ordinary conjugate pairs carry independent uniform phases; nonzero self-conjugate modes carry independent Haar signs. Because the same factor multiplies every component and sampled time,
\begin{equation}
 \widehat u_a^{\phi}(\boldsymbol k,t)
 \widehat u_b^{\phi}(\boldsymbol k,t')^{*}
 =\widehat u_a(\boldsymbol k,t)
 \widehat u_b(\boldsymbol k,t')^{*}.
 \label{eq:quadratic}
\end{equation}
Equation~\eqref{eq:quadratic} preserves the prescribed same-mode quadratic tensor and two-time Gram information. Vector-modal amplitudes, total energy, shell spectra, Hermitian reality, and modewise solenoidality follow. Higher-order and cross-mode statistics remain free to change. In the numerical implementation, the phase action is applied to the admissible reference field; the reference-projection and phase-mask checks are separated in Fig.~\ref{fig:intervention}(a,b).

We filter each velocity component with a normalized spatial kernel:
\begin{equation}
 u_i^\ell(\boldsymbol x,t)=F_\ell[u_i](\boldsymbol x,t)
 =\int_{\mathbb R^3}G_\ell(\boldsymbol r)
 u_i(\boldsymbol x-\boldsymbol r,t)\,d^3r,
 \qquad \int_{\mathbb R^3}G_\ell(\boldsymbol r)\,d^3r=1.
 \label{eq:filter-map}
\end{equation}
The measured flux uses a box average of width $\ell=w\Delta$, where $\Delta$ is the grid spacing. With these positive weights, $\tau_{ij}^\ell$ is a local covariance: the average of $u_i u_j$ minus the product of the two averages. The discrete filter and boundary conventions are given in Supplemental Sec.~\ref{sup:implementation}.

The local flux is
\begin{equation}
 \Pi_\ell=-\tau_{ij}^{\ell}S_{ij}^{\ell},\qquad
 \tau_{ij}^{\ell}=F_\ell[u_i u_j]-u_i^{\ell}u_j^{\ell},\qquad
 S_{ij}^{\ell}=\tfrac12(D_j u_i^{\ell}+D_i u_j^{\ell}),
 \label{eq:flux}
\end{equation}
with positive sign denoting downscale transfer. The phase group contains a central involution that reverses every nonzero-mode fluctuation while leaving the mean velocity fixed. The stress is even under this transformation; the filtered strain is odd. The entire space--time flux array therefore changes sign. Haar invariance gives
\begin{equation}
 \Pi_\ell^{\phi}\overset{\mathrm d}{=}-\Pi_\ell^{\phi},\qquad
 \Ehaar\mathcal O[\Pi_\ell^{\phi}]=0\quad\text{for }\mathcal O[-A]=-\mathcal O[A],
 \label{eq:null}
\end{equation}
under a constant-reproducing filter and a derivative that annihilates constants. For the phase-Haar ensemble, the expectation of any integrable odd flux functional is zero. The result applies jointly to sampled positions and times, so it includes odd observables of finite-window transfer. It is an ensemble identity; a finite unpaired set of masks can retain a nonzero sample residual.

\FloatBarrier

The information boundary is now exact. Two fields related by the central involution share Eq.~\eqref{eq:quadratic} and have opposite signed transfer. The retained quadratic information cannot distinguish them. In Fourier space, the missing information appears in third-order triadic phase combinations. No empirical spectral comparison is needed to establish this boundary; the numerical experiment asks how strongly it is expressed in a finite sample.

\begin{figure}[!htbp]
\centering\includegraphics[width=\textwidth]{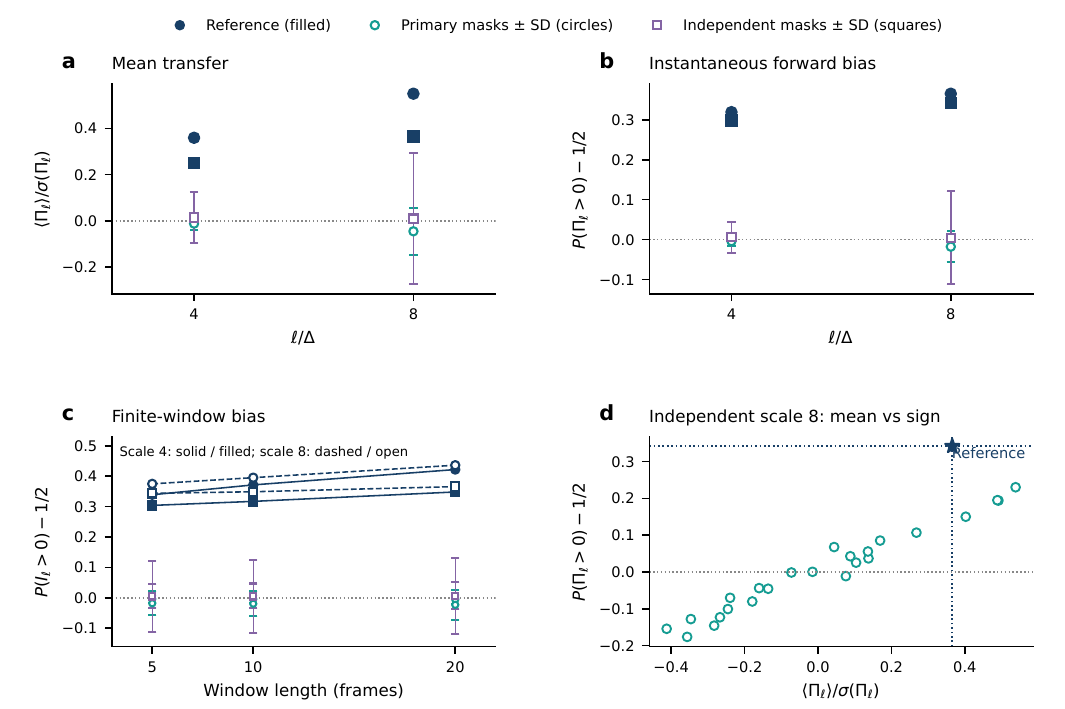}
\caption{Static arrow erasure and its finite-sample scope. (a) Normalized mean flux and (b) instantaneous forward-event bias for the primary (32 masks) and independent (24 masks) blocks. Filled marks denote references; open marks and bars are surrogate means and sample SD. Circles denote the primary block and squares the independent block; teal and purple distinguish their phase ensembles. The independent comparator is the identity-phase projected reference. (c) Accumulated sign bias for 5/10/20 frames with a fixed mask throughout each sequence. (d) All independent scale-8 masks in the normalized mean--sign plane. Sign separation persists, while the prespecified raw-mean criterion fails; four plotted normalized means also exceed the reference.}
\label{fig:static}
\end{figure}

The static test uses forced isotropic turbulence from JHTDB \citep{r51,r52,r53}: a primary $64^3\!\times100$ sequence with 32 masks and a disjoint block with 24 new masks, both at $\ell/\Delta=4$ and 8. We measure the normalized mean $A_\ell=\avg{\Pi_\ell}/\sigma(\Pi_\ell)$ and the forward-event bias $b_\ell=P(\Pi_\ell>0)-1/2$. In the primary block, the reference means are 0.359 and 0.550; both phase ensembles lie near zero. At scale 4, the reference bias is 0.319 and the surrogate mean is $-0.00395$. The separation persists for accumulated transfer over 5, 10, and 20 frames (Fig.~\ref{fig:static}(c)).

The independent block exposes a useful difference between the two measures. Its sign biases remain below the reference, but the scale-8 mean criterion fails: eight of the 24 raw means exceed the projected comparator, and four normalized means exceed it in Fig.~\ref{fig:static}(d). The prespecified criterion requires reference dominance over every mask for raw mean, instantaneous bias, and 10-frame bias at both scales; the other five comparisons satisfy it. Mean and sign need not rank the same phase states alike. The full mask and window results appear in Fig.~\ref{fig:s4}.

\FloatBarrier

The next question is dynamical. Starting from matched phase-scrambled checkpoints in a generated forced periodic $N=64$ flow, we compare full Navier--Stokes evolution (NL), evolution without quadratic advection (LIN), and a frozen-organization control (FRZ). FRZ takes the NL vector-modal amplitude at every output but keeps the initial scrambled normalized complex direction. It supplies an amplitude-matched trajectory without allowing those directions to evolve. Three checkpoints from the same simulated flow, crossed with three preassigned masks, give nine matched cases evaluated with an eight-cell box filter.

For branch $b$, the selected scalar and its recovery are
\begin{equation}
 X_b(t)=\frac{\avg{\Pi_b(t)}_x}{\avg{|\Pi_b(t)|}_x},\qquad
 R_\Pi^{(b)}(t)=1-
 \frac{|X_b(t)-X_{\rm nat}(t)|}{D_{\Pi0}},\qquad
 D_{\Pi0}=|X_{\rm nat}(0)-X_{\rm pert}(0)|>0.
 \label{eq:recovery}
\end{equation}
The initial distance is fixed, and the natural reference evolves. Thus $R=1$ is scalar agreement, $R=0$ is the initial gap, and a negative value is a larger gap.

\begin{figure}[!htbp]
\centering\includegraphics[width=\textwidth]{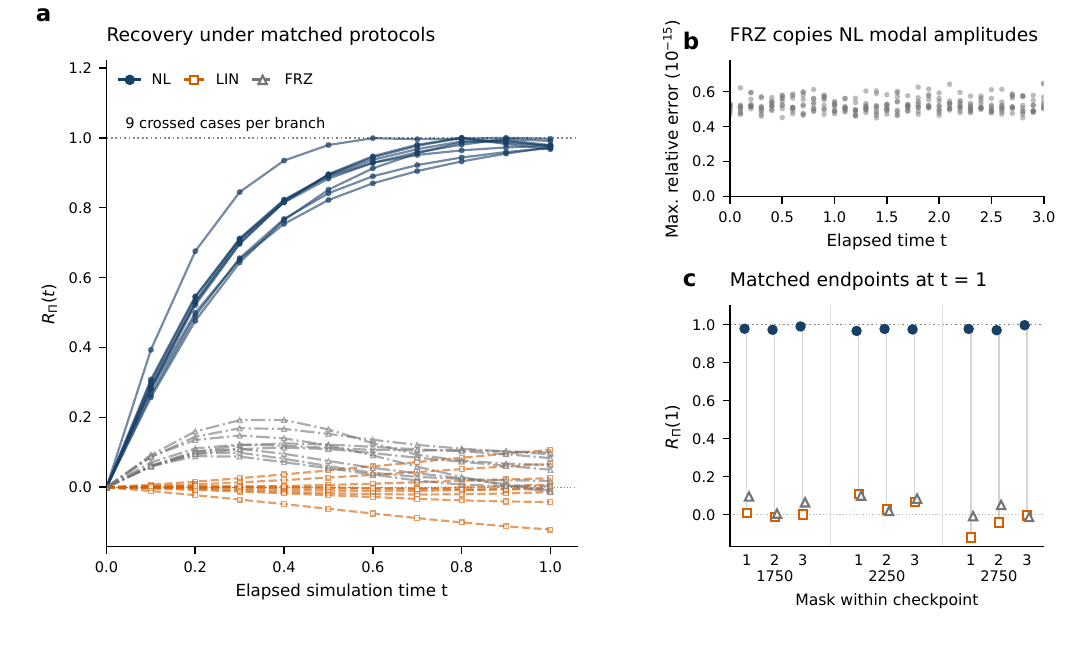}
\caption{Regeneration under matched evolution protocols. (a) Signed-flux recovery for NL, LIN, and constructed FRZ over $0\leq t\leq1$, showing all nine checkpoint--mask cases with eleven recorded times per branch. (b) FRZ--NL relative vector-modal amplitude errors at all 279 outputs over $0\leq t\leq3$. (c) The same cases at $t=1$, grouped by checkpoint (1750/2250/2750), with mask indices 1--3 shown within each group. Small branch offsets are categorical. Recovery is defined by Eq.~\eqref{eq:recovery}; negative values are retained. The generated $N=64$ flow uses an eight-cell box filter. The nine cases form a crossed design, and FRZ is constructed from the NL amplitude trajectory.}
\label{fig:recovery}
\end{figure}

NL recovers rapidly: at $t=1$, all nine cases reach $R_\Pi=0.96688$--$0.99692$ (Fig.~\ref{fig:recovery}a,c).

LIN and FRZ remain far from comparable recovery. Their failure answers two different questions. LIN tests whether the retained viscous and forcing evolution is enough; FRZ asks whether following the nonlinear modal-amplitude history is enough. In the latter, amplitudes track NL to near numerical precision at every recorded output (Fig.~\ref{fig:recovery}b), yet the signed scalar does not follow. The separation is present in every matched endpoint, rather than emerging only after case averaging. Within these protocols, nonlinear evolution restores the selected arrow projection while the two controls do not. Recovery of this scalar is the measured result, rather than reconstruction of the entire field.

\FloatBarrier

\begin{figure}[!htbp]
\centering\includegraphics[width=\textwidth]{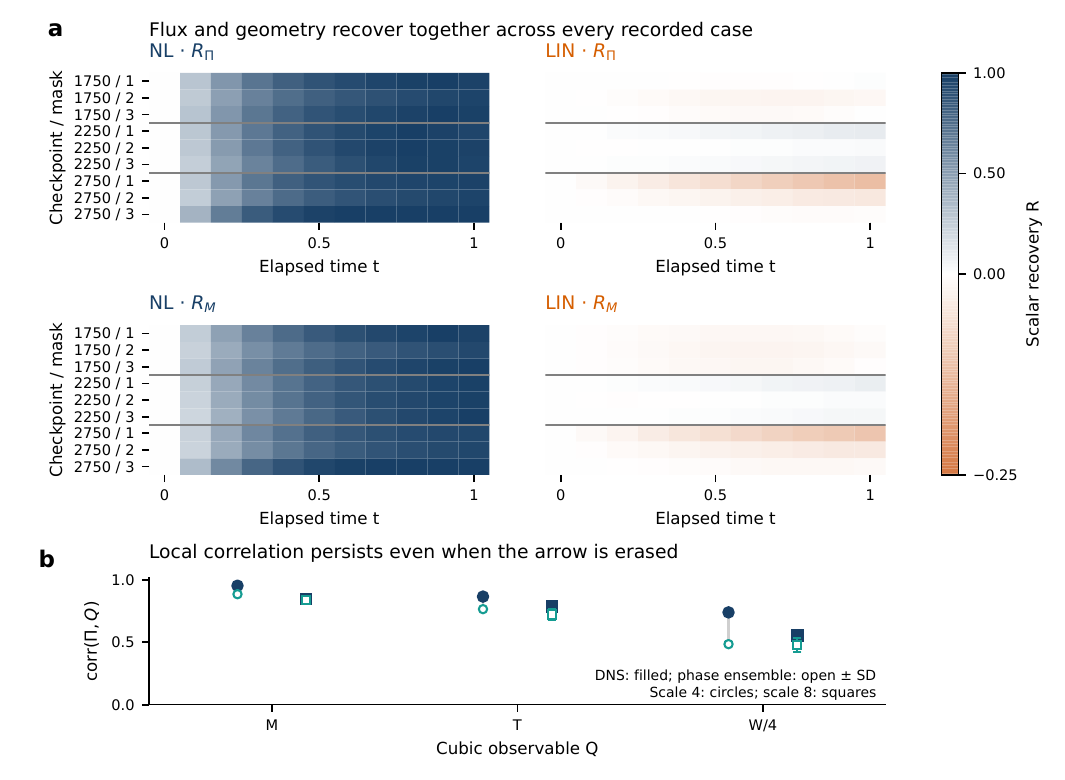}
\caption{Co-recovery and the limits of local association. (a) Flux and $M$ recovery under NL and LIN, using the same case order and color scale. All 396 recorded values over $0\leq t\leq1$ are shown. The added geometry record accompanies the flux recovery of Fig.~\ref{fig:recovery}; FRZ $M$ means were not recorded. (b) Static primary-block correlations of flux with $M,T,W/4$ at scales 4 and 8. Filled marks denote DNS; open marks and bars denote phase-mask means and sample SD. Circles and squares distinguish scales 4 and 8 within each observable category. High local correlation survives arrow erasure. These correlation summaries also appear in Fig.~\ref{fig:s5}(c).}
\label{fig:geometry}
\end{figure}

A local cubic diagnostic supplies a second view of the regeneration:
\begin{equation}
 M=-\mathrm{tr}(S^3)+\tfrac14\omega_iS_{ij}\omega_j,
 \qquad T=-\mathrm{tr}(S^3),\qquad W/4=\tfrac14\omega_iS_{ij}\omega_j.
 \label{eq:geometry}
\end{equation}
Using $Y_b=\avg{M_b}_x/\avg{|M_b|}_x$ in the same fixed-distance construction gives $R_M$ (Supplemental Sec.~\ref{sup:dynamic}). Flux and $M$ recover together under NL; neither shows comparable LIN recovery (Fig.~\ref{fig:geometry}a). This comparison concerns the recorded NL and LIN geometry means.

The static correlation tells a different story. At scale 8, the pooled flux--$M$ correlation is about 0.845 in DNS and 0.837 in the phase ensemble, even though the signed bias has been strongly suppressed (Fig.~\ref{fig:geometry}b). A field can therefore retain a strong local association between flux and cubic geometry while losing its global directional preference.

Local correlation alone is not sufficient to fix the cascade arrow.

\FloatBarrier

The dynamic contrast has a simple symmetry interpretation. A center-equivariant evolution carries a center-symmetric initial law into another center-symmetric law. Quadratic advection has a different parity,
\begin{equation}
 B(-u,-u)=B(u,u),
\end{equation}
and can generate an odd response. That response is allowed by symmetry, rather than guaranteed by the presence of a nonlinear term. The NL/LIN comparison shows it in the tested forced flow; FRZ then separates the recorded amplitude trajectory from the evolving organization that carries the flux. Details of the state-dependent forcing law and matching are given in the Supplement.

This interpretation places the result alongside phase-dynamics studies \citep{r30,r31,r33,r34}, but gives the comparison a different starting point. Equation~\eqref{eq:quadratic} holds fixed the modal tensor and two-time record before the arrow is tested. The theorem identifies the information absent from that record; the recovery experiment asks how it reappears. Modal occupation and directed exchange are thus separated at the level of the construction, not merely by resemblance between spectra.

The geometry result adds a further distinction. Strain self-amplification, vortex stretching, and their multiscale interactions are central to cascade descriptions \citep{r60,r61,r62,r63,r65,r66}. The leading gradient expansion motivates the cubic combination $M$ \citep{r14,r15}, while the velocity-gradient literature emphasizes the organization of strain and vorticity \citep{r67,r68,r69,r70,r71,r72,r73}. Here the two measurements answer different questions: NL co-recovery links the return of a signed flux scalar to the return of $M$; the static intervention shows that their local correlation can persist without a directional population bias. Taken together, they identify a limit on sufficiency, rather than a unique causal role for geometry.

The scale of regeneration remains open. A natural-state $K/\varepsilon$ candidate increases the observed across-case spread by about 11.1\% instead of improving collapse (Fig.~\ref{fig:s6}a,b). With the adopted spread measure, it offers no improvement over native time. This negative result is useful precisely because the curves recover similarly at native time: a shared trend alone does not establish a universal clock.

\begin{samepage}
We have separated where turbulent energy resides from how it is transferred. A Hermitian phase intervention preserves a prescribed quadratic record while erasing odd transfer in ensemble expectation. The finite data display that separation, including an independent mean failure, and the matched nonlinear dynamics regenerates the selected signed scalar together with a local cubic diagnostic. The cascade arrow is therefore carried by organization beyond the retained same-mode information---organization that the tested nonlinear evolution can rebuild.

\end{samepage}
\FloatBarrier
\section*{Acknowledgments}
The Johns Hopkins Turbulence Database and its supporting institutions are acknowledged.
\section*{Code and data availability}
The source isotropic-turbulence fields are publicly available from JHTDB \citep{r53}. Numerical code, processed data, phase-mask identifiers, initial Fourier checkpoints, and figure-generation scripts are available at \url{https://github.com/zhaoyang7719286-hub/turbulent-cascade-arrow}. The repository provides workflows for reconstructing Main Figs.~1--4 and Supplemental Figs.~S1--S6 from the supplied records and for running the matched NL/LIN/FRZ cases from the supplied checkpoints. Numerical definitions, matching protocols, and implementation checks are given in the Supplemental Material.

\FloatBarrier
\clearpage
\input{supplement.tex}
\clearpage
\input{main.bbl}

\end{document}

%% file: authors.tex
\begin{center}
Yang Zhao\textsuperscript{1}, Daocheng Zhou\textsuperscript{1}, Hongfu Zhang\textsuperscript{2}\\[4pt]
{\small\textsuperscript{1}College of Civil Engineering and Transportation, Northeast Forestry University, Harbin, China\\
\textsuperscript{2}Department of Mechanical Engineering, The Hong Kong Polytechnic University, Hong Kong, China}
\end{center}

%% file: supplement.tex
\setcounter{section}{0}
\setcounter{equation}{0}
\setcounter{figure}{0}
\renewcommand{\thesection}{S\arabic{section}}
\renewcommand{\theequation}{S\arabic{equation}}
\renewcommand{\thefigure}{S\arabic{figure}}
\renewcommand{\theHsection}{sup.\arabic{section}}
\renewcommand{\theHequation}{sup.\arabic{equation}}
\renewcommand{\theHfigure}{sup.\arabic{figure}}
\begin{center}
{\Large\bfseries Supplemental Material}\\[5pt]
{\large Erasing and Regenerating the Turbulent Cascade Arrow}
\end{center}

\section{Coarse-grained energy balance and parity}
Filtering the incompressible Navier--Stokes equation gives
\begin{equation}
 \partial_t\frac{|\boldsymbol u^\ell|^2}{2}
 +\nabla\cdot\boldsymbol J_\ell
 =-\Pi_\ell-\nu|\nabla\boldsymbol u^\ell|^2
 +\boldsymbol u^\ell\cdot\boldsymbol f^\ell,
\end{equation}
where $\boldsymbol J_\ell$ collects resolved, pressure, viscous, and subgrid transport. Positive $\Pi_\ell$ removes energy from the resolved scales.

Center inversion reverses the nonzero-mode fluctuation $\boldsymbol v(\boldsymbol x,t)$ at fixed sampled times. Kinematic time reversal instead maps $\boldsymbol u(\boldsymbol x,t)$ to $-\boldsymbol u(\boldsymbol x,-t)$, and dynamical time-reversal symmetry additionally requires a transformed trajectory to satisfy the same equation. The phase-null theorem uses the first transformation. The cascade arrow refers to signed interscale transfer and its accumulation, rather than thermodynamic entropy production.

\begin{figure}[!htbp]
\centering\includegraphics[width=\textwidth]{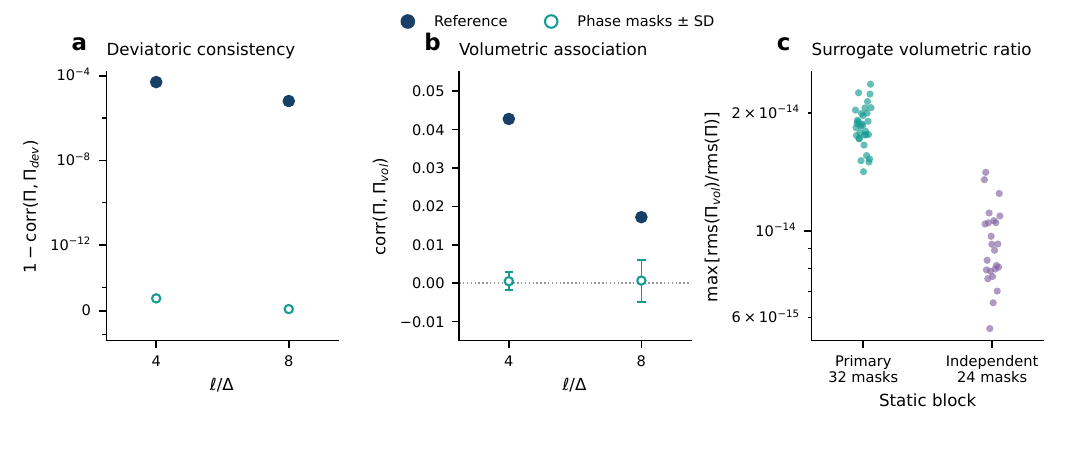}
\caption{Deviatoric and volumetric contributions to the corrected flux. (a) $1-\mathrm{corr}(\Pi,\Pi_{\rm dev})$ and (b) $\mathrm{corr}(\Pi,\Pi_{\rm vol})$ at scales 4 and 8. Filled marks denote DNS; open marks and bars denote primary-mask means and sample SD. The symlog axis in (a) retains zero and roundoff signs. (c) Per-mask maximum RMS ratios $\mathrm{rms}(\Pi_{\rm vol})/\mathrm{rms}(\Pi)$ for all 32 primary and 24 independent masks. Each point in (c) is one mask; horizontal offsets only separate points within a block.}
\label{fig:s1}
\end{figure}

Using the measured residual divergence $\theta=D_i u_i^\ell$,
\begin{equation}
 \Pi_\ell=\Pi_{\rm dev}+\Pi_{\rm vol},\qquad
 \Pi_{\rm dev}=-\tau_{ij}^\ell\left(S_{ij}^{\ell}-\tfrac13\delta_{ij}\theta\right),
 \qquad\Pi_{\rm vol}=-\tfrac13\theta\tau_{ii}^\ell.
\end{equation}
The corrected flux closely follows the deviatoric term, while the volumetric contribution has a small RMS amplitude (Fig.~\ref{fig:s1}). These summaries assess the size of the residual-divergence contribution.

\FloatBarrier

\section{Phase sampler, projection, and admissibility}\label{sup:implementation}
In a nonredundant real Fourier basis, equal-variance independent Gaussian cosine and sine coefficients give circular complex coefficients. Their phases are uniform on $U(1)$. Nonzero self-conjugate modes supply independent Haar signs on $\mathbb Z_2$, and the zero-mode factor is set to unity. One sampled factor is used for all components and frames.

The fourth-order centered derivative in the static discrete calculation has symbol
\begin{equation}
 \kappa_j(\boldsymbol k)=\frac{8\sin(k_j\Delta)-\sin(2k_j\Delta)}{6\Delta}.
\end{equation}
For $|\boldsymbol\kappa|>0$, the discrete Helmholtz projector is
\begin{equation}
 P_{D,ij}(\boldsymbol k)=\delta_{ij}-\frac{\kappa_i\kappa_j}{|\boldsymbol\kappa|^2},\qquad
 P_D(\zeta_{\boldsymbol k}\widehat{\boldsymbol u})
 =\zeta_{\boldsymbol k}P_D\widehat{\boldsymbol u}.
\end{equation}
The norm-restoring scalar depends on phase-independent pre- and post-projection norms and also commutes with the action. Derivative-null modes follow the fixed numerical convention. Equation~\eqref{eq:quadratic} therefore applies to the admissible field entering the intervention. The independent comparator, \texttt{reference\_star32}, is the identity-phase output of this same projection and rescaling.

\subsection*{Spatial filtering}
The continuous box kernel is
\begin{equation}
 G_\ell(\boldsymbol r)=\ell^{-3}
 \prod_{j=1}^3\mathbf 1_{\{|r_j|\leq\ell/2\}},
 \label{eq:box-kernel}
\end{equation}
where $\mathbf 1$ is the indicator function. The numerical box average uses the even widths $w=4,8$ in the static tests and $w=8$ in the dynamic comparison. For a grid index $\boldsymbol m$, its exact discrete form is
\begin{equation}
 F_{w\Delta}^{\rm disc}[f](\boldsymbol m)
 =\frac{1}{w^3}\sum_{\boldsymbol r\in\mathcal W_w}
 f_{\rm ext}(\boldsymbol m+\boldsymbol r),
 \qquad
 \mathcal W_w=\{-w/2,\ldots,w/2-1\}^3.
 \label{eq:discrete-box}
\end{equation}
The velocity components and their products use the same weights. Static blocks extend the nearest boundary value outside the sampled cube; the dynamic periodic flow wraps indices modulo $N$. These are the \texttt{nearest} and \texttt{wrap} modes, respectively, of \texttt{scipy.ndimage.uniform\_filter} with its default origin. The static D4 derivative uses periodic index rolls, whereas the dynamic gradients are spectral. The Gaussian Fourier filter of Sec.~\ref{sup:triads} is used only for the generated identity audits.

\subsection*{Shell energy and normalization}
For a sampled cube of side $L_b=N\Delta$, write the Fourier wave vector as $\boldsymbol k=2\pi\boldsymbol n/L_b$. With normalized coefficients,
\begin{equation}
 \widehat u_i(\boldsymbol n,t)=\frac{1}{N^3}
 \sum_{\boldsymbol m}u_i(\boldsymbol m,t)
 e^{-2\pi\mathrm i\boldsymbol n\cdot\boldsymbol m/N},
 \qquad
 E_q(t)=\frac12\sum_{\boldsymbol n\in\mathcal S_q}
 \sum_{i=1}^3|\widehat u_i(\boldsymbol n,t)|^2.
 \label{eq:shell-energy}
\end{equation}
The supplied static phase-gate routine uses $\mathcal S_q=\{\boldsymbol n:q\leq|\boldsymbol n|<q+1\}$, or $q=\lfloor|\boldsymbol n|\rfloor$. Thus each shell collects modes with similar wave-number magnitude. The zero shell contains the spatial mean; nonzero modes have wavelength $L_b/|\boldsymbol n|$, so larger $q$ denotes smaller spatial scales. Shell index $q$ and filter width $w=\ell/\Delta$ are different coordinates.

Parseval's identity gives $\sum_qE_q=\tfrac12\langle|\boldsymbol u|^2\rangle_x$. The static phase-gate routine stores spectra from unnormalized FFT coefficients, giving $N^6E_q$; this common factor cancels in its energy ratios and relative spectral errors. Fig.~\ref{fig:intervention}(a) is reproduced from the deposited mean-shell diagnostic table, using its recorded shell indices and common normalization
\begin{equation}
 E_0=\sum_{q\geq0}\overline E_q^{\rm DNS},
 \qquad \text{plotted ordinate}=\overline E_q/E_0\quad(q>0).
 \label{eq:spectrum-normalization}
\end{equation}
Here $\overline E_q$ denotes the recorded mean shell value. The normalization includes the zero shell for every curve; $E_0$ is the reference total, not the energy of shell zero. Unit-modulus scalar phases preserve every modal energy, and therefore preserve shell sums for any fixed partition of modes.

\begin{figure}[!htbp]
\centering\includegraphics[width=\textwidth]{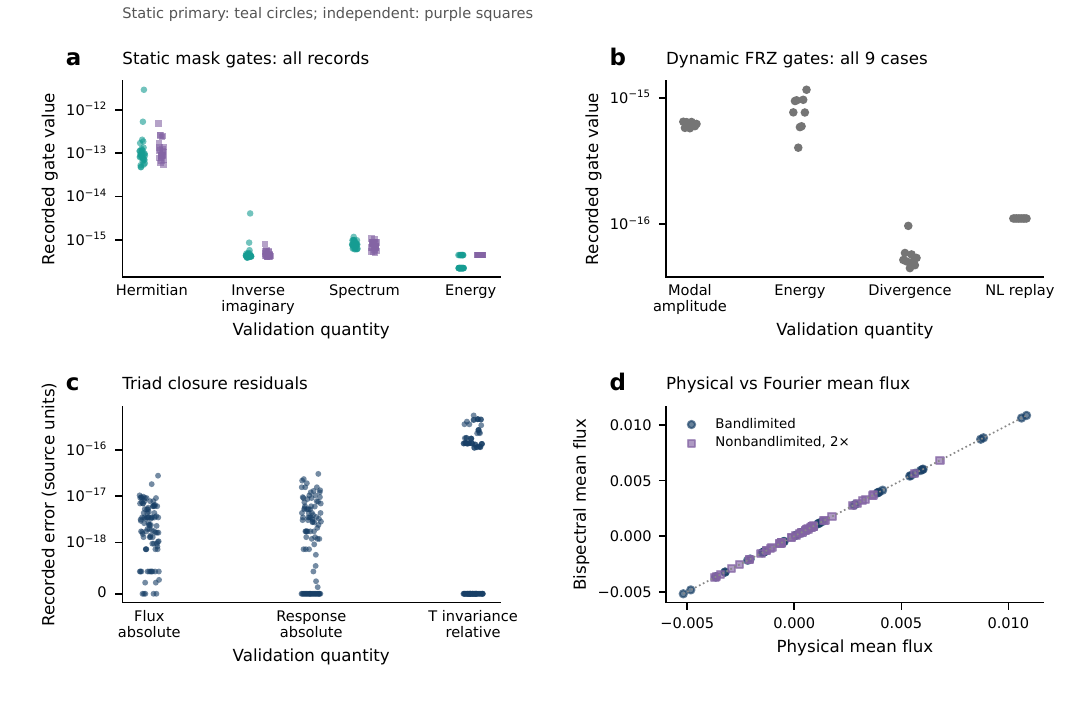}
\caption{Numerical admissibility and flux identities. (a) Static checks for all masks in the two $64^3$ blocks. (b) FRZ amplitude, energy, divergence, and NL replay checks for the nine dynamic cases. (c) Flux closure, phase-response closure, and triadic-capacity invariance residuals in generated $16^3$ fields; the panel's $T$ is $\mathcal T$, not the cubic gradient variable. (d) Physical-space and bispectral mean flux in all 96 tests, using spectral/D4 derivatives and filter widths 2/4. The line is identity. In (a--c), the horizontal axis lists validation quantities; narrow offsets separate records and carry no numerical scale. Symlog axes retain zero.}
\label{fig:s2}
\end{figure}

Figure~\ref{fig:s2}(a) reports reality, inverse-transform imaginary part, spectrum, and energy checks for both $64^3$ static blocks. Dynamic checks add FRZ amplitude matching, energy, divergence, and NL replay consistency (b). The identity tests in (c,d) use generated $16^3$ fields. In these tests the label $T$ denotes the nonnegative triadic capacity $\mathcal T$ of Sec.~\ref{sup:triads}, distinct from the gradient scalar $T=-\mathrm{tr}(S^3)$.

\FloatBarrier

\section{Proof of the exact phase-null theorem}\label{sup:proof}
Write $\boldsymbol u=\boldsymbol c+\boldsymbol v$, with fixed zero mode $\boldsymbol c$. For a constant-reproducing filter,
\begin{equation}
 \tau_{ij}^\ell[\boldsymbol c+\boldsymbol v]
 =F_\ell(v_i v_j)-F_\ell(v_i)F_\ell(v_j)
 =\tau_{ij}^\ell[\boldsymbol c-\boldsymbol v].
\end{equation}
A derivative that annihilates constants gives
\begin{equation}
 S_{ij}^\ell[\boldsymbol c-\boldsymbol v]
 =-S_{ij}^\ell[\boldsymbol c+\boldsymbol v].
\end{equation}
The full flux array consequently changes sign, pointwise at every sampled position and time.

Let $C$ multiply every nonzero $U(1)$ or $\mathbb Z_2$ factor by $-1$. This central element preserves product Haar measure. For an integrable odd functional of the complete array,
\begin{equation}
 \Ehaar\mathcal O[\Pi]=\Ehaar\mathcal O[\Pi\circ C]
 =\Ehaar\mathcal O[-\Pi]=-\Ehaar\mathcal O[\Pi]=0.
\end{equation}
The proof needs no assumption about independent flux values. A functional may combine multiple positions and times, including an accumulated transfer over a chosen window. Using a fixed mask is what keeps the entire temporal record under one group action.

The exact odd sign statistic is $\tfrac12\avg{\mathrm{sgn}\Pi}$, with $\mathrm{sgn}(0)=0$. It agrees with the reported $P(\Pi>0)-1/2$ when there is no mass at zero; otherwise the zero-event contribution distinguishes them. Finite unpaired masks need not cancel, whereas each mask and its antithetic counterpart cancel every odd statistic exactly \citep{r50}.

\section{Fourier triadic identity and audit}\label{sup:triads}
Under the normalized periodic Fourier convention, multiplication becomes modular convolution. For a real symmetric filter $G_{\boldsymbol k}$ and odd derivative symbol $\boldsymbol\kappa$, the ordered-triad expression is
\begin{align}
 J_\ell(\zeta)&=\sum_{\boldsymbol k+\boldsymbol p+\boldsymbol q=0}
 W_{\boldsymbol k\boldsymbol p\boldsymbol q}
 \mathrm{Im}\!\left[C_{\boldsymbol k\boldsymbol p\boldsymbol q}
 \zeta_{\boldsymbol k}\zeta_{\boldsymbol p}\zeta_{\boldsymbol q}\right],\\
 W_{\boldsymbol k\boldsymbol p\boldsymbol q}
 &=G_{\boldsymbol k}(G_{\boldsymbol k}-G_{\boldsymbol p}G_{\boldsymbol q}),\qquad
 C_{\boldsymbol k\boldsymbol p\boldsymbol q}
 =\kappa_j(\boldsymbol k)\widehat u_i(\boldsymbol k)
 \widehat u_i(\boldsymbol p)\widehat u_j(\boldsymbol q).
\end{align}
Wave-vector closure is modulo the grid. The imaginary part comes from differentiation, and the filter difference comes from filtering a product versus multiplying filtered fields. All conventions are shared with the physical-space calculation.

The generated tests use $N=16$ on $[0,2\pi]^3$, normalized FFT coefficients, and a Gaussian transfer $G_{\boldsymbol k}=\exp[-\sigma^2|\boldsymbol k|^2/2]$ with $\sigma=w\Delta/\sqrt{12}$ for widths $w=2,4$. Bandlimited fields retain $|k_j|\leq3$; the nonbandlimited tests remove Nyquist planes and evaluate products on a $32^3$ padded grid. Spectral and D4 derivatives are compared in 96 configurations (Fig.~\ref{fig:s2}c,d).

The phase-independent capacity is
\begin{equation}
 \mathcal T=\sum_{\boldsymbol k+\boldsymbol p+\boldsymbol q=0,i,j}
 |W_{\boldsymbol k\boldsymbol p\boldsymbol q}\kappa_j(\boldsymbol k)|
 |\widehat u_i(\boldsymbol k)\widehat u_i(\boldsymbol p)\widehat u_j(\boldsymbol q)|,
 \qquad |J_\ell|\leq\mathcal T.
\end{equation}
For $\mathcal T>0$, $C_\phi=J_\ell/\mathcal T$ measures signed organization relative to this capacity. Scalar phases retain $\mathcal T$ while changing the triadic phase factors. Zero-mode coefficients vanish through constant reproduction or the zero derivative. Nonzero self-conjugate phases average as independent signs. These cases are all covered by the group proof above.

\section{Finite-window statistics}
Accumulated transfer uses consecutive frames with one fixed mask across the sequence. We report windows of 5, 10, and 20 frames. Independent frame-by-frame masks would define a different intervention and remove the retained same-mode temporal record.

\begin{figure}[!htbp]
\centering\includegraphics[width=\textwidth]{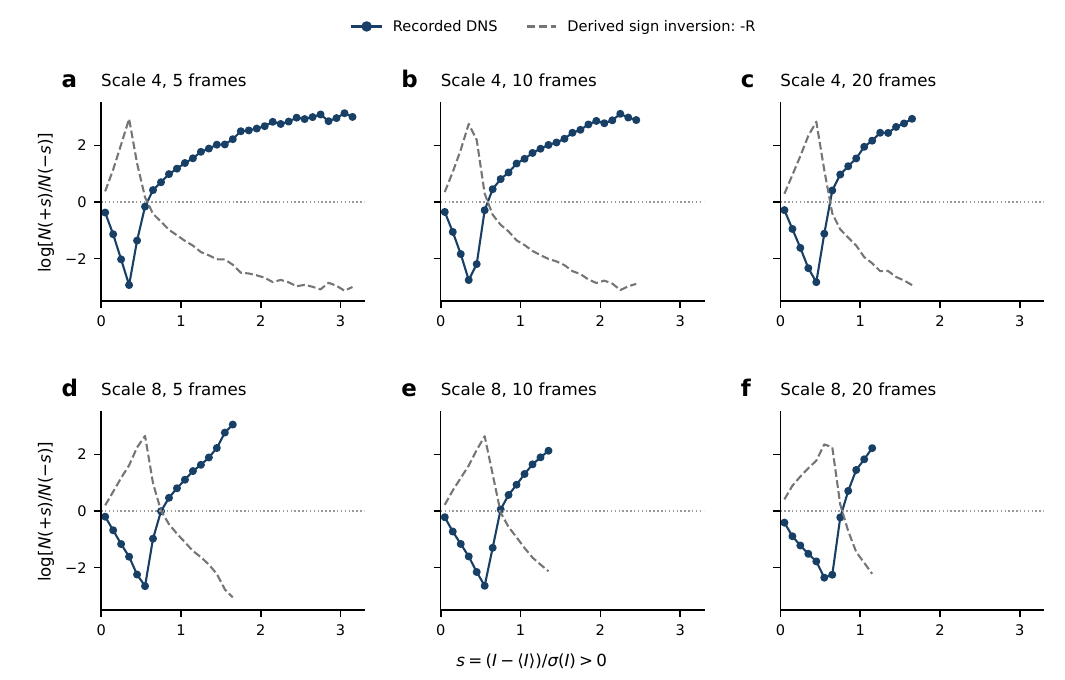}
\caption{Centered finite-window DNS asymmetry. Rows correspond to scales 4 and 8; columns to 5, 10, and 20 frames. Solid curves join the recorded supported bins, each with at least 50 counts in both tails. Dashed curves are the algebraically derived sign-inverted counterpart, $-\mathcal R$, obtained by exchanging those counts. They are not empirical phase-null curves.}
\label{fig:s3}
\end{figure}

For the centered DNS distribution diagnostic in Fig.~\ref{fig:s3},
\begin{equation}
 s=\frac{I-\avg I}{\sigma(I)}>0,\qquad
 \mathcal R(s)=\ln\frac{N(+s)}{N(-s)}.
\end{equation}
Each paired tail must contain at least 50 counts. The 117 supported centered bins are retained. Swapping the same counts gives the algebraic sign-inverted counterpart $-\mathcal R$; the dashed curves are not a measured phase ensemble. Longer windows change the supported interval, so the curves are interpreted only within their sampled tails.

\FloatBarrier

\section{Extended static ensemble results}
\begin{figure}[!htbp]
\centering\includegraphics[width=\textwidth]{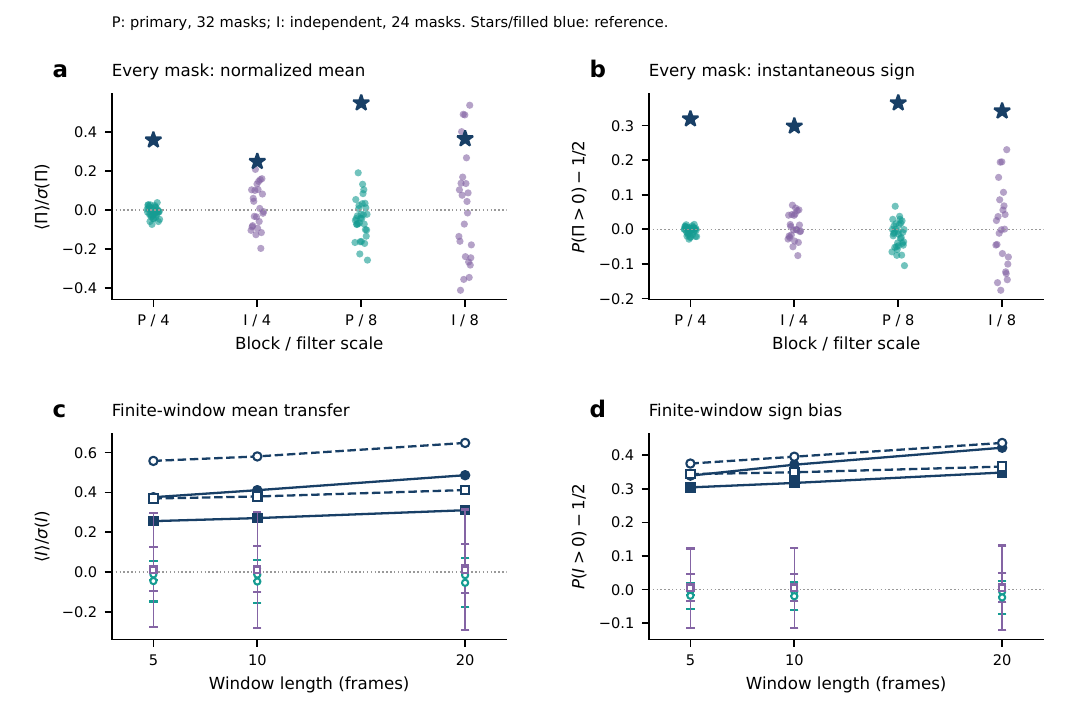}
\caption{Static scope across masks and windows. (a,b) All 112 mask--scale normalized means and instantaneous biases. Stars denote the DNS primary or projected independent reference. The horizontal categories specify block and filter scale; within-category offsets separate masks. (c,d) Accumulated normalized means and biases for 5/10/20 frames, with mask sample SD; solid and dashed lines denote scales 4 and 8. The independent scale-8 mean failure remains visible, alongside sign separation. Both blocks have $64^3$ spatial resolution.}
\label{fig:s4}
\end{figure}

The mask-level comparisons retain 112 values per metric: two scales for 32 primary and 24 independent masks (Fig.~\ref{fig:s4}a,b). The prespecified independent criterion requires the projected reference to strictly exceed all 24 surrogates for raw mean flux, instantaneous bias, and 10-frame bias at both scales. The scale-8 raw mean fails this requirement, with eight exceedances; the other five comparisons pass. Four scale-8 normalized means also exceed the reference. Sign biases for every recorded mask stay below their comparator.

Panels (c,d) extend the comparison across all three windows. The independent means remain more dispersed, while the reference sign bias stays above the phase-ensemble center. Panel (d) provides the full window summary used in main Fig.~\ref{fig:static}(c).

\FloatBarrier

\section{Dynamic protocols and matched controls}\label{sup:dynamic}
Forcing choices affect DNS statistics \citep{r56}. We use a periodic spectral discretization \citep{r57,r58} with $N=64$, viscosity $\nu=0.015$, step $\Delta t=0.01$, and constant-power input 0.1 through negative damping on wave-number magnitudes 1--2. NL and LIN use the same forcing law and viscosity; LIN omits quadratic advection. The force is state dependent. Checkpoints 1750, 2250, and 2750 are crossed with masks 2026074101, 2026074102, and 2026074103. All branches start from the same projected, dealiased perturbed field. Flux uses an eight-cell box filter; outputs are spaced by 0.1 over $0\leq t\leq3$.

For active nonzero modes,
\begin{equation}
 \widehat{\boldsymbol u}_{\rm FRZ}(\boldsymbol k,t)
 =\|\widehat{\boldsymbol u}_{\rm NL}(\boldsymbol k,t)\|
 \frac{\widehat{\boldsymbol u}_{\rm scr}(\boldsymbol k,0)}
 {\|\widehat{\boldsymbol u}_{\rm scr}(\boldsymbol k,0)\|}.
\end{equation}
FRZ copies the vector-modal norm at each output and holds the initial normalized complex direction fixed. Its maximum relative amplitude error across 279 outputs is $6.47\times10^{-16}$ (Fig.~\ref{fig:s2}b). It is a constructed trajectory, not an autonomous solver branch.

Flux recovery uses main Eq.~\eqref{eq:recovery}. For geometry, write
\begin{equation}
 Y_b(t)=\frac{\avg{M_b(t)}_x}{\avg{|M_b(t)|}_x},\qquad
 R_M^{(b)}(t)=1-\frac{|Y_b(t)-Y_{\rm nat}(t)|}{D_{M0}},\qquad
 D_{M0}=|Y_{\rm nat}(0)-Y_{\rm pert}(0)|>0.
\end{equation}
Both distances are fixed initially while the natural references evolve. The nine checkpoint--mask combinations are a crossed design. FRZ $M$ means are absent, so the geometry comparison uses NL and LIN.

\section{Local-gradient diagnostics}
\begin{figure}[!htbp]
\centering\includegraphics[width=\textwidth]{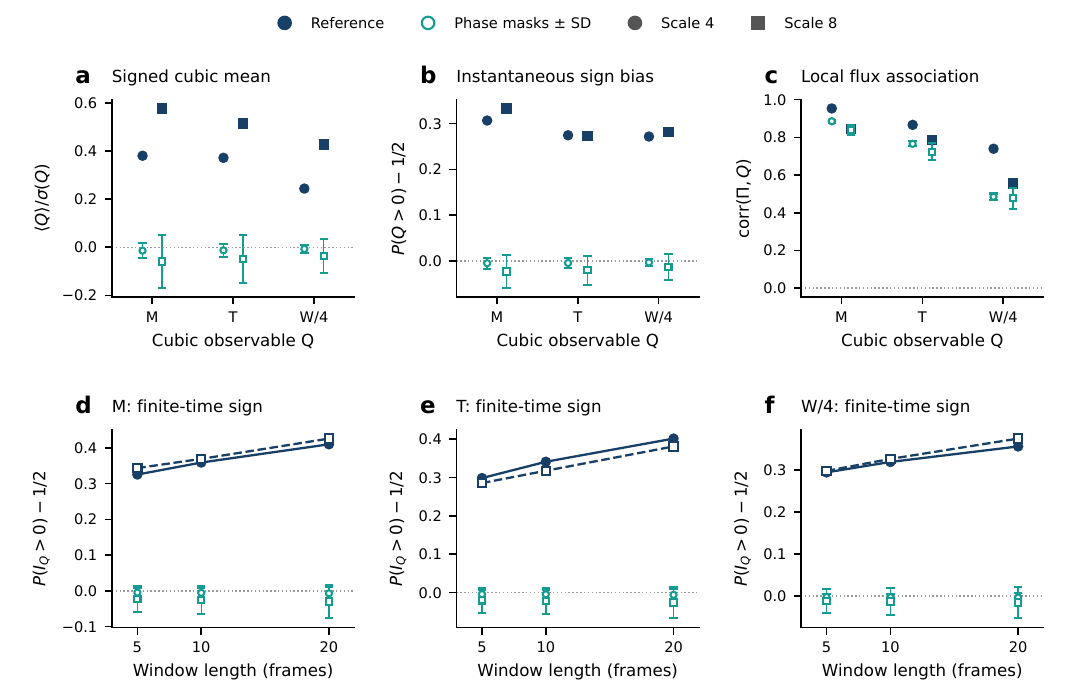}
\caption{Local cubic diagnostics. (a) Normalized signed means, (b) instantaneous forward-event biases, and (c) pooled flux correlations for $M,T,W/4$ at scales 4 and 8. (d--f) Their accumulated biases over 5/10/20 frames; solid and dashed lines denote scales 4 and 8. Filled marks denote DNS; open marks and bars denote phase-mask means and sample SD. Circles and squares denote scales 4 and 8 within each observable category. Panel (c) retains the full diagnostic summary of main Fig.~\ref{fig:geometry}(b).}
\label{fig:s5}
\end{figure}

The leading gradient expansion for a smooth symmetric filter is
\begin{equation}
 \tau_{ij}^\ell\simeq C_G\ell^2\partial_k u_i^\ell\partial_k u_j^\ell,
 \qquad
 \Pi_\ell^{\rm Clark}\simeq C_G\ell^2
 \left[-\mathrm{tr}(S^3)+\tfrac14\omega_i S_{ij}\omega_j\right].
\end{equation}
It motivates the local cubic variable $M$ rather than an exact closure of the measured flux. Figure~\ref{fig:s5} compares $M,T,W/4$ at both static scales: means and instantaneous biases in (a,b), pooled correlations in (c), and accumulated biases in (d--f). The signed statistics weaken under the phase ensemble while positive local correlations persist. Panel (c) is the correlation record interpreted in main Fig.~\ref{fig:geometry}(b).

\FloatBarrier

\section{Time normalization and the full recovery interval}
\begin{figure}[!htbp]
\centering\includegraphics[width=\textwidth]{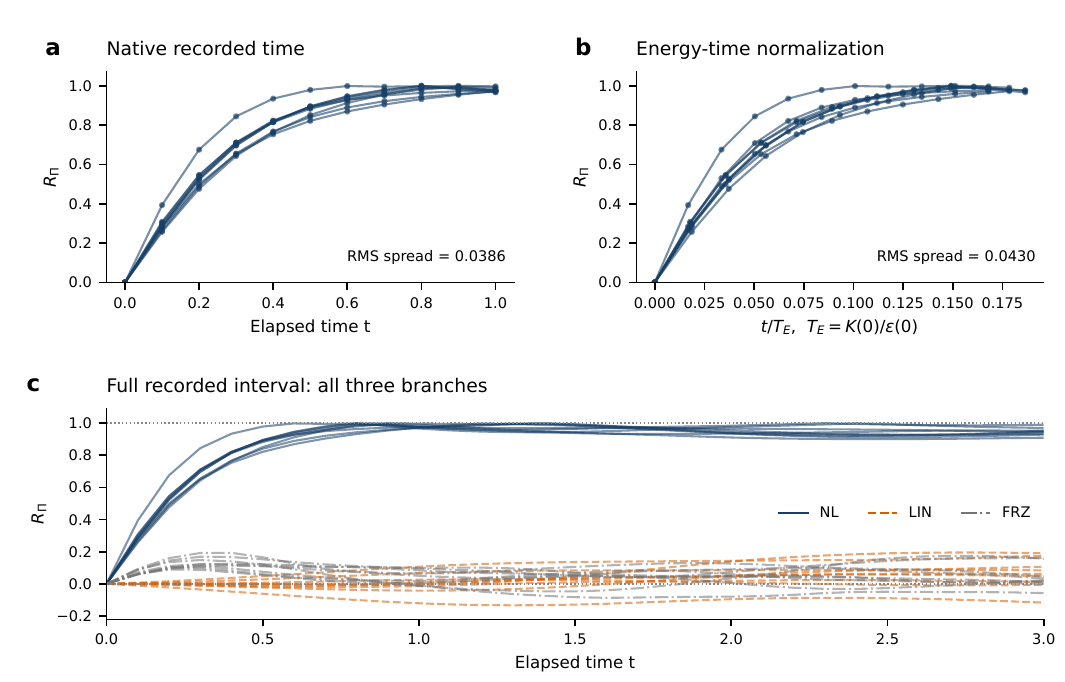}
\caption{Candidate time scale and extended records. (a,b) The same NL recovery curves at native time and $t/T_E$, with $T_E=K_{\rm nat}(0)/\varepsilon_{\rm nat}(0)$. The spread increases by 11.1\%; this normalization does not improve collapse. (c) All NL/LIN/FRZ records through $t=3$. The cases are those of main Fig.~\ref{fig:recovery}.}
\label{fig:s6}
\end{figure}

The unfitted candidate time is $T_E=K_{\rm nat}(0)/\varepsilon_{\rm nat}(0)$, with checkpoint values 5.60355, 5.36278, and 5.95142. Figure~\ref{fig:s6}(a,b) compares the same nine NL curves at native time and $t/T_E$. The descriptive RMS across-case spread rises from 0.03864 to 0.04295, an increase of about 11.1\%. According to this measure, the candidate normalization does not improve collapse.

Piecewise-linear alignment on common supported early-time coordinates is used only to calculate that spread. Displayed points remain the recorded observations. Panel (c) extends the branch comparison to $t=3$ and retains negative values and late-time variability.

\FloatBarrier
\section{Computational records and reproduction}\label{sup:reproduction}
The accompanying repository contains the numerical kernels, phase-mask identifiers, three initial Fourier checkpoints, and processed records used in the figures. The figure workflow reconstructs Main Figs.~1--4 and Supplemental Figs.~S1--S6 from these records, retaining the mask, scale, window, and checkpoint identifiers. It runs offline after dependency installation. Dynamic workflows evolve the matched NL/LIN/FRZ cases from the supplied checkpoints using the protocol of Sec.~\ref{sup:dynamic}.

The preparation environment uses Python~3.12 with pinned dependencies. The repository README gives the installation and execution commands; its validation documentation records the numerical and figure checks performed. The deposited static data are processed records, while the original source fields are obtained from JHTDB. Data provenance and upstream acquisition dependencies are documented alongside the code.
\FloatBarrier